# Stable crack propagation in dislocation-engineered oxide visualized by double cleavage drilled compression test

Oliver Preuß[1a], Zhangtao Li[2a], Enrico Bruder[3], Philippe Carrez[4], Yinan Cui[2*], Jürgen Rödel[1], Xufei Fang[5*]

[1]Division Nonmetallic-Inorganic Materials, Department of Materials and Earth Sciences, Technical University of Darmstadt, Peter-Grünberg-Str. 2, 64287 Darmstadt, Germany

[2]Department of Engineering Mechanics, Tsinghua University, Mengminwei Building, Beijing 100084, P.R. China

[3]Division Physical Metallurgy, Department of Materials and Earth Sciences, Technical University of Darmstadt, Peter-Grünberg-Str. 2, 64287 Darmstadt, Germany

[4]Université Lille, CNRS, INRAE, Centrale Lille, UMR 8207—UMET—Unité Matériaux et Transformations, Lille, France

[5]Institute of Applied Materials, Karlsruhe Institute of Technology, Kaiserstr. 12, 76131 Karlsruhe, Germany

[a] These authors contributed equally to this work.

* Corresponding authors: Yinan Cui: cyn@mail.tsinghua.edu.cn; Xufei Fang: xufei.fang@kit.edu

**Abstract:**

Understanding crack tip – dislocation interaction is critical for improving the fracture resistance of semi-brittle materials like room-temperature plastically deformable ceramics. Here, we use a modified double cleavage drilled compression (DCDC) specimen geometry, which facilitates stable crack propagation, to achieve *in situ* observation of crack tip – dislocation interaction. MgO specimens, furnished with dislocation-rich barriers, were employed to study how dislocations influence crack propagation. Crack progression was clearly observed to decelerate within dislocation-rich regions, slowing to 15% of its velocity as compared to the pristine crystal. Upon exiting these regions, cracks reaccelerated until reaching the next dislocation-rich barrier. Coupled phase field and crystal plasticity modeling replicates the experimental observations and provides mechanistic insight into crack tip – dislocation interactions. The aligned experiment and simulation results underscore the robustness of the technique and its potential to inform the design of more fracture-resistant ceramics via dislocations.

# 1. Introduction

The ability to resist crack growth via dislocations has led to the wide application of engineering metals and alloys as structural materials. Phenomenologically, in these materials, the plastic deformation at the crack tip involves dislocation emission, multiplication, and motion. This contributes to the shielding and/or blunting of the crack tip, which inhibits and/or delays crack propagation, leading to a desirable engineering property for damage-tolerant components. In ceramics, however, crack propagation is energetically favoured over plastic deformation, which significantly limits their application in structural components compared to metals [1]. In recent years, the promising development in dislocation-engineered ceramics together with more discovered plastically deformable ceramics [2, 3] suggests the possibility of dislocation-based toughening in ceramics even at room temperature. This is evidenced by the experimental observations in oxides and alkali halides [2, 4-9] that an area of high-density dislocations can reduce the length of surface cracks induced by Vickers indentation, or completely suppress crack formation. However, direct observation of the crack tip – dislocation at mesoscale is still missing, primarily due to the challenge of engineering sufficiently large plastic zones and achieving stable crack growth in ceramics. To enable the development of dislocation-toughened ceramics, it is important to pinpoint the underlying mechanisms by directly observing crack propagation in oxides with dislocations in a controlled manner.

For both metals and ceramics, several crack tip – dislocation interaction mechanisms have been proposed to describe the origin of the dislocation toughening effect. This includes, for instance, dislocation shielding (lowering of the local stress intensity factor by the presence of a suitably oriented dislocation stress field) [4, 10]; dislocation emission (the crack front acts as a heterogeneous dislocation source) [5, 11]; dislocation-based crack-tip blunting (an avalanche of emission events, which leads to accumulating surface steps on the crack front changing the crack tip geometry) [12, 13]; and crack-tip plasticity (while propagating, the crack-tip stress field acts as a Peach-Köhler force on pre-existing dislocations, effectively performing mechanical work) [14, 15]. Furthermore, pile-up events of dislocations, such as the Zener-Stroh mechanism [16-18] or Cottrell mechanism [19], can result in local stress concentration and eventually crack initiation. Provided that the cracks are initiated, the core challenge for dislocation toughening shifts to crack tip – dislocation interaction during crack propagation.

Experimentally, a direct, *in situ* method capable of tracking dislocations in the volume of interest much larger than the process zone in a time-resolved manner is still missing for ceramics. Such a method has to be developed to identify the prevalent mechanism in dislocation-based toughening for future exploitation. Here, we summarize the most promising experimental approaches along this direction in **Table 1**.

*Table 1. Comparison of possible experimental approaches for observing crack tip – dislocation interactions.*

|  |  | *In situ* observation | Dislocation resolution | Large volume of interest/ area of interest | Challenges and limitations |
|---|---|---|---|---|---|
| **Macroscale (> 1 mm)** | Notched bend bars [4, 20] | Yes | No | Yes | high sample cost for ceramic single-crystals |
| **Mesoscale (10 µm – 1 mm)** | Indentation / Electron channelling contrast imaging [2, 6-9] | No | Yes | Yes | Concurrent crack initiation and propagation |
|  | Double cleavage drilled compression/Electron channelling contrast imaging | Yes | Possibly | Yes |  |
|  | Dark-field X-ray microscopy | Possible | Yes | Yes | Requires adequate beam line with compatible mechanical loading setup |
| **Microscale (<10µm)** | Micromechanical fracture geometries [21, 22] | Yes | Possibly | No | FIB damage, high image forces |
|  | Transmission electron microscopy [5, 22, 23] | Yes | Yes | No | FIB damage, high image forces, potential electron beam effect |

One of the most appealing approaches listed in **Table 1** is the *in situ* transmission electron microscopy (TEM) observation by Appel et al. on single-crystal MgO at room temperature [23]. At a very slow crack propagation rate (the source suggests a velocity of micrometers per hour), the formation of a dislocation-rich "plastic zone" (in the form of a process zone) in front of a loaded crack tip was observed. The dislocations were generated by profuse multiplication in front of the crack tip. Based on the configuration of the dislocations, the local stress intensity factor can be estimated to be ~0.55 MPa m$^{0.5}$. A similar method was used later by Higashida et al. [5], who observed spontaneous emission of mainly shielding-type screw dislocations from a mode-I crack tip in single-crystal MgO. However, because of the thin TEM sample (in the range of hundreds of nm), the stress state during *in situ* TEM experiments differs from those encountered in bulk samples, necessitating caution in the extrapolation to bulk samples [24].

For ceramic materials, most testing geometries listed in **Table 1** with notches and pre-cracks (to circumvent crack initiation) obey the Griffith criterion [25]. This implies a rising driving force with increasing crack length, which leads to unstable crack growth. When a crack reaches the propagation criterion ($K_I \geq K_{IC}$) in brittle solids, it can accelerate to the speed of sound [26] or until the sample breaks entirely, making the detailed observation impossible. Therefore, for detailed evolutionary characterisation in brittle materials, stable crack growth with $\frac{\partial K_I}{\partial a} \leq 0$ is necessary, as the crack may be stopped at designated time/loading.

To achieve stable crack propagation in brittle solids, the double cleavage drilled compression (DCDC) geometry was first used by Janssen in 1974 (compare **Fig. 1A**) [27]. This specimen geometry, tested in a widely available compression setup, offers an inverse proportionality of stress intensity factor and crack length ($K_I \propto \frac{1}{a}, \frac{\partial K_I}{\partial a} < 0$) [28]. This means that, for a given increase in applied stress, the crack will extend a finite length and theoretically stop by itself in the loaded state without the need for significant unloading. In other words, the crack growth in this experimental configuration is self-limiting. With a step-wise increase in applied load, the crack length will increase step-wise. Further, the DCDC geometry offers midplane crack stability and self-precracking [29]. The scalability of the DCDC test is mostly limited in brittle materials by the precise machining processes involved in the sample manufacturing, mainly the achievable hole and notch sizes. The DCDC geometry also enables fatigue testing in very brittle samples [30]. It is mostly used as a quantitative method to determine the fracture toughness of brittle materials but also used to *in situ* observe crack deflection [31] and crack front shapes [32, 33] when combined with X-ray methods or digital image correlation method.

Here, we adopt the DCDC test on single-crystal MgO to demonstrate stable crack propagation and capture its interaction with the dislocation-rich areas at the mesoscale, for the first time. MgO is chosen as a representative oxide that exhibits room-temperature dislocation plasticity in bulk compression [34-36]. The test is carried out inside an SEM chamber to allow for concurrent *in situ* observation of crack morphology and dislocation activity. Combined with electron channelling contrast imaging (ECCI), new insight into crack tip – dislocation interaction in *ductile* oxides has been obtained. Furthermore, a coupled crystal plasticity and phase field modelling is carried out for the DCDC test with a focus on the crack tip – dislocation interactions to corroborate the experimental findings.

## 2. Experimental procedure

### 2.1. Sample preparation

MgO single crystals (2 x 4 x 8 mm$^3$) with well-polished surfaces were purchased from Alineason Materials Technology GmbH (Frankfurt am Main, Germany) for the double cleavage drilled compression (DCDC) geometry given in **Fig. 1A**. All 8 mm long edges were chamfered to a 45° angle using P4000 sandpaper and a fixture to hold the crystal at this angle to prevent stress concentrations and edge flaws. The hole in the center was drilled using a water-cooled diamond hollow core drill (Günther Diamantwerkzeuge e.K., Idar-Oberstein, Germany) with an outer diameter of 1.2 mm on a conventional milling machine.

The pre-cracks were prepared using diamond wire that was threaded through the hole and tensioned in a jigsaw. The opposing notches of ~0.5 mm depth were cut by hand with a sprayed water droplet to facilitate the cutting. The sawed notches were later refined using a sharp razor blade and 1 µm diamond paste. For reproducibility, three samples were prepared and tested under the same conditions. Each sample contains 6 scratch tracks (orange strips in **Fig. 1A**), yielding in total 18 tracks furnished with dislocations, acting as barriers right in front of the crack propagation path.

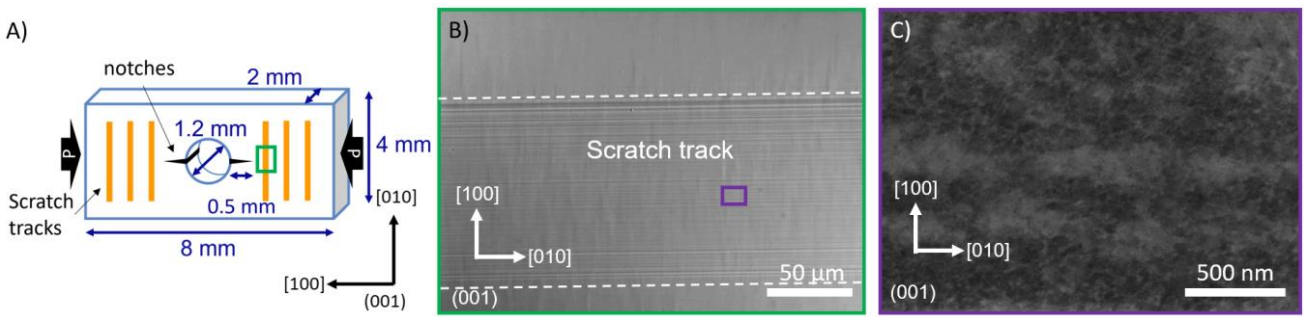

*Figure 1. A) Double cleavage drilled compression (DCDC) sample geometry. The orange rectangles symbolize the positions of the dislocation-rich scratch tracks. B) Differential Interference Contrast imaging of a single scratch track (between the two white dashed lines) highlighting the slip traces. C) Electron channeling contrast imaging (ECCI) inside the scratch track as marked by the purple box in B, revealing the high-density dislocations (white contrasts).*

The dislocation-rich regions (i.e. the scratch tracks) were achieved using the cyclic Brinell scratching method developed elsewhere [37]. A hardened steel ball of 2.5 mm diameter (Habu Hauck Prüftechnik GmbH, Hochdorf-Assenheim, Germany) was used with a static load of 9.8 N (1 kgf) and a velocity of 0.5 mm/s on a modified universal hardness tester (Finotest, Karl Frank GmbH, Weinheim, Germany). The lateral movement was implemented by a spindle-driven stage (Thorlabs Inc., Newton, NJ, USA) controlled by a custom LabVIEW programme. While the pristine MgO single crystals feature a dislocation density of ~10$^{12}$ m$^{-2}$, the dislocation density increases with the number of passes during scratching, with a saturation observed after about 10 passes [9]. Based on our previous work, to achieve a high dislocation density (~10$^{14}$ - 10$^{15}$ m$^{-2}$) in MgO to maximize the crack-arresting effect, 10 passes were repeated within each scratch track. The final scratch tracks have a width of ~120 µm, a length of 3.5 mm, and are 0.5 mm apart from each other. It has been reported that the dislocations in such scratch tracks on MgO penetrate as deep as the width of the scratch track due to the inherent slip systems of the ½<110>{110} type being activated [9]. These slip planes either lie at an angle of 45° or 90° to the scratched surface. Introducing scratch tracks like this can lead to significant residual stresses [9], which might influence the cracking behaviour.

The successful introduction of dislocations was confirmed by differential interference contrast (DIC) imaging using an optical microscope (AxioImager 2, Carl Zeiss Microscopy GmbH, Jena, Germany) and by electron channelling contrast imaging (ECCI) [38] inside a scanning electron microscope (SEM, MIRA3-XM, Tescan, Brno, Czech Republic). In the optical microscope images, the slip traces are visualized as vertical and horizontal lines in and around the scratch track (**Fig. 1B**). Employing ECCI, the individual dislocations in the wear track appear as small white tails of high contrast (see **Fig. 1C**), which allows estimation of dislocation density. Here, the mean distance between dislocations in the black areas is roughly 50 nm, which translates into an estimated dislocation density of $4 \times 10^{14}$ m$^{-2}$ [9], which is 2 orders of magnitude higher than that in the surrounding reference material.

**2.2. Mechanical testing**

The compressive loading of the DCDC specimen was performed with a small spindle-driven load frame (SEMtester, MTII/Fullam, Albany, NY, USA) placed inside the SEM. To handle the inevitable transverse contraction of the sample during loading, which could lead to stress concentrations and premature failure, copper sheets (1 mm thickness) were placed on either end of the sample as a malleable interlayer. This also contributes to the compliance of the loading system. Surface charging effects during SEM observation were mitigated by a ~10 nm carbon layer on the top surface of the DCDC sample, allowing for better imaging quality. Silver paste was used to connect the carbon layer with the copper sheet electrically, thus the load frame, which was grounded.

After the sample was clamped in the correct orientation in the load frame using a preload of 10 N, the load frame was loaded into the SEM. The compressive load was increased in steps of 50 N until a fresh crack emerged from the pre-notch. Afterwards, 10 N and 1 N increments were used to drive the crack to propagate in a controlled manner. Despite the very stable behaviour of the crack, the load was reduced by 10 N while images were being made. This prevents any further crack propagation during imaging [39], which can take up to 5 minutes.

**3. Modelling**

**3.1. A coupled crystal plasticity and phase field model**

A coupled crystal plasticity and phase-field model (CP-PFM) is developed to capture the interplay between plasticity and fracture for MgO. The crystal plasticity module describes dislocation mechanism-controlled plastic deformation through incorporating the atomistic scale informed dislocation mobility law and the mesoscale-informed dislocation density evolution equation. The phase-field model represents the evolution of discrete crack topology using a continuous damage field. The specific details of CP-PFM have been provided in our previous paper [40]. Here, we will only introduce the fundamental equations and the developments made to adapt the method for MgO.

Through using the continuous damage field variable *d*, it is easy to capture complicated crack topology. When *d* = 0, the material remains intact, whereas *d* = 1 indicates the crack is fully formed. The governing equation of damage variable is derived according to the first and second laws of thermodynamics [40-42] and is given by **Eqs. (1-3)**:

$$\frac{G_c}{l_0}(d - l_0^2 \Delta d) = 2(1-d)(\psi_+^{e0} + \psi^p) \tag{1}$$

$$\psi_+^{e0} = \frac{\lambda}{2}\langle \varepsilon_1^e + \varepsilon_2^e + \varepsilon_3^e \rangle_+^2 + \mu(\langle \varepsilon_1^e \rangle_+^2 + \langle \varepsilon_2^e \rangle_+^2 + \langle \varepsilon_3^e \rangle_+^2) \tag{2}$$

$$\psi^p = (1-\chi)\int_0^{\varepsilon^p} \boldsymbol{\sigma}:d\boldsymbol{\varepsilon}^p \tag{3}$$

Here, $G_c$ denotes the critical energy release rate, and $l_0$ represents the characteristic length that governs the width of the diffusive crack. As $l_0$ approaches zero, the diffusive crack formulation converges to the discrete crack. $\Delta$ is the Laplace operator defined as $\Delta = \sum_i^3 \frac{\partial^2}{\partial x_i^2}$ in three-dimensional (3D) space. The term $\psi_+^{e0}$ is the tensile part of the elastic strain energy, where $\lambda$ and $\mu$ are referred to Lamé's first constant and shear modulus for the intact material, respectively. Since compressive stress on the crack surface does not drive crack propagation, the compressive component does not contribute to damage evolution. $\psi^p$ represents the inherent stored energy density due to plasticity, determined by the Taylor-Quinney coefficient $\chi$. $\chi$ is the fraction of plastic work converted to heat, generally taken as 0.9 [43, 44]. $\varepsilon_1^e, \varepsilon_2^e, \varepsilon_3^e$ are the first, second, and third principle elastic strains, respectively. $\boldsymbol{\varepsilon}^p$ is the plastic strain tensor, and $\boldsymbol{\sigma}$ is the stress tensor. In the presence of a damage field, $\boldsymbol{\sigma}$ is reduced from its value in the intact state, as described by the following equation:

$$\boldsymbol{\sigma} = g(d)\boldsymbol{C}:(\boldsymbol{\varepsilon} - \boldsymbol{\varepsilon}^p) \tag{4}$$

where $\boldsymbol{C}$ is the fourth-order elastic modulus tensor of the intact material, and $\boldsymbol{\varepsilon}$ is the strain tensor. $g(d)$ is the degradation function, generally taken as $g(d) = (1-d)^2$ [45].

The plastic deformation, after dislocation engineering, is dominated by the dislocation motion in the current experimental case. The plastic strain rate tensor $\dot{\boldsymbol{\varepsilon}}^p$ is expressed as:

$$\dot{\boldsymbol{\varepsilon}}^p = \frac{1}{2}\sum_\beta \dot{\gamma}^\beta(\boldsymbol{m}^\beta \otimes \boldsymbol{n}^\beta + \boldsymbol{n}^\beta \otimes \boldsymbol{m}^\beta), \tag{5}$$

where $\dot{\gamma}^\beta$ indicates the slip rate on the $\beta$-th slip system defined by the plane normal $\boldsymbol{n}^\beta$ and slip direction $\boldsymbol{m}^\beta$ (**Eq. 4**). Since the activation of ½<110>{100} slip systems in MgO requires very high critical shear stresses, in the following, we only consider the activation of ½<110>{110} slip systems at room temperature. The slip rate can be calculated from the Orowan relation (**Eq. 6**):

$$\dot{\gamma}^\beta = b\rho^\beta v^\beta \tag{6}$$

where $b$ is the magnitude of the Burgers vector, $\rho^\beta$ and $v^\beta$ are the mobile dislocation density and dislocation velocity in the $\beta$-th slip system, respectively.

In MgO, the motion of screw dislocations is primarily governed by the kink-pair mechanism, with the atomic simulation-informed dislocation mobility laws provided in **Eq. 7** [46-48]:

$$v = b\frac{L}{l_c}\nu_D\frac{b}{l_c}\exp\left(\frac{-\Delta H(\tau_{eff})}{k_B T}\right) \tag{7}$$

where $l_c$ is the critical width of kink pairs, $L$ represents the length of dislocation segment, $\nu_D$ is the Debye frequency, $k_B$ is the Boltzmann constant and $T$ is the absolute temperature. $\Delta H(\tau_{eff})$ is the activation enthalpy of kink-pair nucleation as a function of the effective resolved shear stress $\tau_{eff}$ acting on the considered dislocation by:

$$\Delta H(\tau_{eff}) = \Delta H_0\left[1 - \left(\frac{\tau_{eff}}{\tau_P}\right)^p\right]^q, \tag{8}$$

where $\tau_P$ is the Peierls stress, $p$ and $q$ are fitting parameters from atomic information. $\Delta H_0$ is the enthalpy barrier for kink nucleation. As suggested by a previous study [46], the mobility law for edge dislocations is considered to follow the same framework as that of screw dislocations, but with a velocity ten times that of screw dislocations.

The effective shear stress is calculated by:

$$\tau_{eff}^\beta = \boldsymbol{\sigma}:(\boldsymbol{m}^\beta \otimes \boldsymbol{n}^\beta) - \alpha_d\mu b\sqrt{\sum_{\beta=1}^{NAS}\rho^\beta}, \tag{9}$$

where $\boldsymbol{\sigma}$ is the stress tensor, $\alpha_d\mu b\sqrt{\sum_{\beta=1}^{NAS}\rho^\beta}$ considers the dislocation forest hardening and is related to the number of active slip systems NAS, and $\alpha_d$ is the hardening coefficient, $\mu$ is the shear modulus. The dislocation density evolution is described by [49-51]:

$$\dot{\rho}^\beta = \left(k_1\sqrt{\rho^\beta} - k_2\rho^\beta\right)\dot{\gamma}^\beta \tag{10}$$

where $k_1$, $k_2$ are the generation and annihilation coefficients of the dislocations, respectively.

### 3.2. Simulation settings

In order to use fine meshes while also optimizing computation efficiency to capture the physical picture, a two-dimensional (2D) model is initially employed to examine the influence of introducing high dislocation density region, as shown in **Fig. 2**. Symmetric boundary conditions are applied to the left side of the model, with the blue zone representing the high dislocation density region of width of 150 μm, according to experiments. A stress boundary condition is imposed on the right-hand side of the simulation box, and the displacement in the thickness direction is constrained to prevent instability and buckling. A linearly increasing stress, from 0 to 300 MPa, was applied over 1 ms. To maintain a quasi-static loading condition consistent with experiment, the kinetic and internal energies were tracked, and the simulation proceeded only when the kinetic energy was confirmed to be less than 1% of the internal energy. The dimensions of the sample are $W$ = 4 mm and $H$ = 2 mm, respectively. The hole radius is $R$ = 0.5 mm, and the initial crack length is $a$ = 0.5 mm. A global mesh size of 40 μm is employed, with local refinement to 5 μm in the vicinity of the crack tip and along the potential crack propagation path. The analysis utilizes C3D8R elements (8-node linear brick elements with reduced integration) within the ABAQUS framework, resulting in a total of 110,000 elements in the computational domain.

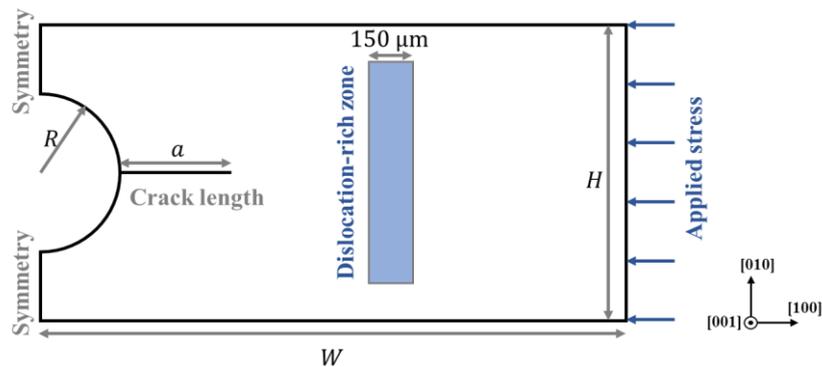

*Figure 2. Geometry of the simulation settings. The dimensions of the sample are W = 4 mm and H = 2 mm, respectively. The hole radius is R = 0.5 mm, and the initial crack length is a = 0.5 mm.*

The parameters used in this study are summarized in **Table 2**. Six ½<110>{110} slip systems for room temperature are considered in the simulations. Within the high dislocation density zone, each slip system is populated with a dislocation density of $4 \times 10^{11}$ m$^{-2}$, with equal densities of screw and edge dislocations, while the rest of the sample remains dislocation-free.

*Table 2. Relevant parameters of MgO.*

| Type | Symbol | Meaning | Value | Unit | Ref. |
|---|---|---|---|---|---|
| **Material** | $\mu$ | Shear modulus | 131 | GPa | [52] |
| | $\nu$ | Poisson's ratio | 0.18 | - | [53] |
| | $G_c$ | Critical energy release rate of {100}-plane | 4.96 | N/m | [54] |

| Dislocation mobility 1/2<110>{110} | $\tau_p$ | Peierls stress | 150 | MPa | [46] |
|---|---|---|---|---|---|
| | $\Delta H_0$ | Critical enthalpy for kink pair nucleation | 1.14 | eV | [46] |
| | $p$ | Fitting parameters | 0.5 | - | [46] |
| | $q$ | Fitting parameters | 2 | - | [46] |
| | $l_c$ | Critical width of kink-pairs | 33.8 | nm | [46] |

## 4. Results and Analyses

### 4.1. Mechanical testing

**Figure 3** demonstrates a representative sequence of images with increasing external stress, following the evolution of a crack tip. The applied external stress was calculated by dividing the applied force by the cross-sectional area of the DCDC specimen (7.2 mm²). Before the stress reaches 100.0 MPa, no change occurred. At 100.0 MPa, a crack is initiated from the notch, which propagates and is stopped by the first barrier **(Fig. 3A)**. At 102.8 MPa, the crack starts to propagate again but now within a dislocation-rich zone. In this region, the crack propagates at a lower velocity, where the crack length increases only slightly for each increment in applied load. As the load continues to rise to 107.0 MPa, the crack eventually exits the dislocation-rich zone and reaches its rear border **(Fig. 3B)**. At this point, the resistance drops significantly, and the crack accelerates, moving quickly through the material. However, its progress is soon halted by a second barrier, which imposes a similar resistance to that of the first, stopping the crack's growth once more and even branching the crack tip **(Fig. 3C)**. This process of stopping, slow propagation within the dislocation-rich zone, and accelerating upon exit of the crack tip is repeated at every dislocation barrier **(Fig. 3D-F)** While the crack is perfectly straight in between these barriers **(Fig. 3D)**, which are near or in the dislocation-rich zones, bridging and deflection can be observed **(Fig. 3A, C, F)**. More extreme cases of bridging and deflection can be seen in the supplementary materials (**Fig. S1**).

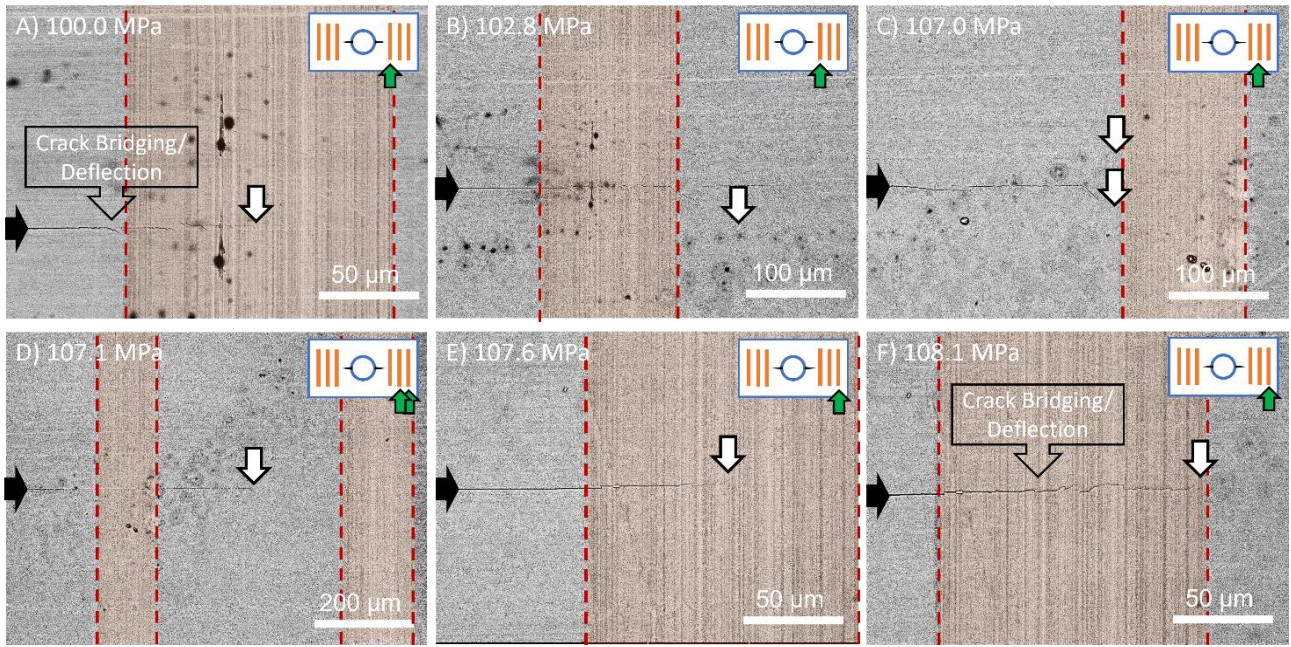

*Figure 3. Sequential loading "snapshots" of DCDC specimen with dislocation-rich barriers (between red dashed lines). The schematic in the top right corner depicts the position on the sample according to Figure 1. The location of the crack front with respect to each of the plastic zone (orange color shaded strips) is indicated by the green arrows. The white arrows mark the position of the crack tip within each subfigure.*

### 4.2. Simulation results

### 4.2.1. Stress field analysis during propagation

The simulation results on crack propagation under applied load are illustrated in **Fig. 4**, where the contour plot of the maximum principal stress is presented. As the load increases, the stress field at the crack tip gradually intensifies. The crack starts to propagate when the applied stress reaches 90 MPa, similar to the experiment. At time $t$ = 0.60 ms, when the crack tip encounters the dislocation-rich region, the crack-tip stress field is shielded by the plastic deformation induced by dislocation motion, leading to a reduction in the local stress intensity factor, as evidenced by the stress contour (color code change) at the crack tip in **Fig. 4A-B**. This leads to a significant decrease in crack propagation velocity, which subsequently stabilizes within the dislocation zone (see **Supplementary Video 1** for the animation). By $t$ = 0.63 ms, the crack propagation velocity decreases to approximately 15% of its value outside the dislocation zone. At $t_4$ = 0.86 ms, as the crack tip exits the dislocation zone, the crack accelerates, resuming to a state similar to that observed in the absence of dislocations. This matches the experimental observation described in the previous section, despite not reaching a complete stop of the crack tip. Furthermore, the simulated DCDC sample endures stresses in excess of 200 MPa, while the real one fails already at 110 MPa. This can be attributed to the kinetic effects of the much higher loading rate in the modelling approach. The presence of crystal defects in the real crystal (e.g. vacancies, impurities, grown-in dislocations, etc.), which was not considered in the simulation, may also add to this discrepancy.

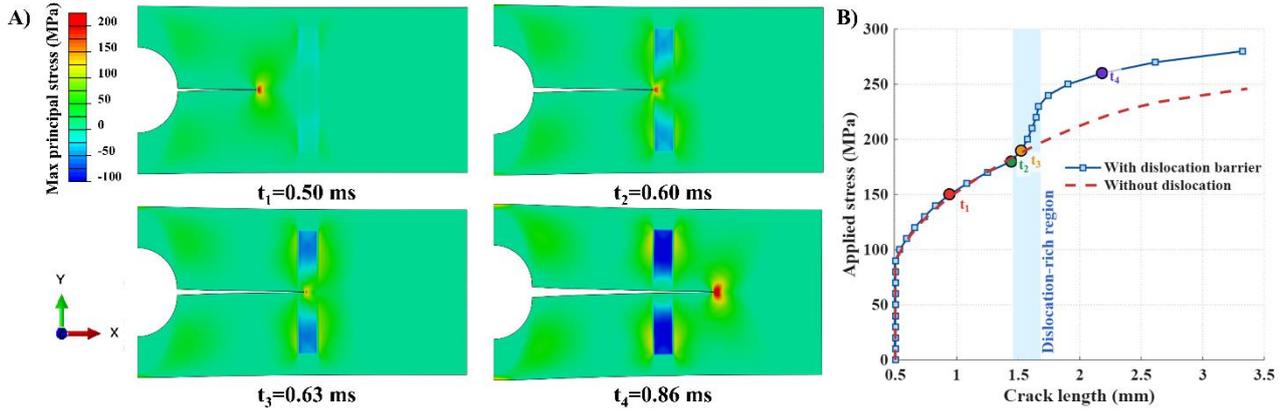

*Figure 4. A) sequential loading "snapshots" of DCDC specimen in CPPFM simulation representing the maximum principal stress. t represents time. B) Plot of applied stress and the resultant crack length. The definition of crack length is consistent with Figure 2.* The deformation scale factor is 10.

**Figure. 4B** presents a comparative analysis of crack propagation with and without the dislocation barrier. Note that a compressive stress is observed near the boundary of the dislocation zone. From this figure, it can be concluded that the dislocation zone effectively reduces the crack propagation velocity. However, when the crack tip is located outside the dislocation zone, its propagation is minimally influenced by the dislocations.

To better understand the dislocation behavior in the dislocation zone upon crack interaction, **Fig. 5** presents the evolution of dislocation density for three representative slip systems: ½$[110](\bar{1}10)$, ½$[01\bar{1}](011)$ and ½$[101](\bar{1}01)$, during the process of the crack tip traversing the dislocation-rich region. Results for the other three activated slip systems are not depicted, as they are equivalent to the these three presented. The dislocation-density increase rate in ½$[101](\bar{1}01)$ slip system (**Fig. 5C**) is an order of magnitude lower than that in the slip system ½$[110](\bar{1}10)$ (**Fig. 5A**), demonstrating that plastic deformation is predominantly accommodated by dislocation glide on the ½$[101](\bar{1}01)$ slip system. The ½$[01\bar{1}](011)$ slip system is active exclusively near the crack tip, characterized by the highest dislocation-density increase rate, as illustrated in **Fig. 5B**.

Two primary dislocation-active regions are identified: the vicinity of the crack tip and the dislocation boundary zone. Along the 45°-inclined plane (the $(\bar{1}10)$ plane) intersecting the crack front, dislocation glide is initiated by the elevated resolved shear stress components generated by the crack-tip stress field. Upon the crack-tip impinging on the dislocation-rich region, the dislocation-density increase rate reaches its maximum. Subsequently, due to the shielding effect of these emitted dislocations, the amplitude of the crack-tip stress field is reduced (**Fig. 4A**). Concurrently, the increasing dislocation density induces forest hardening, requiring a higher resolved shear stress to drive further dislocation motion. The combined influence of the reduced crack-tip stress and the increased resistance consequently suppresses dislocation activity eventually (**Fig. 5A$_2$ and 5A$_3$**).

At the periphery of the dislocation-rich region, the material undergoes a transition from elastic to plastic deformation. This behavior can be attributed to the inherent limitations in dislocation nucleation mechanisms within MgO crystals under initial dislocation-free conditions. The stress of homogeneous dislocation nucleation in MgO can be estimated by $\mu/10 \approx 13$ GPa ($\mu$ is the shear modulus) [55], which is two orders of magnitude than the stress required for dislocation motion (~40 MPa in average) in MgO at room temperature [56]. Under this condition, the dislocation can only multiply and move in the presence of engineered dislocations zones prior to the crack propagation. This leads to the heterogeneous deformation characteristics at the crack tip when intersecting with the dislocation zone.

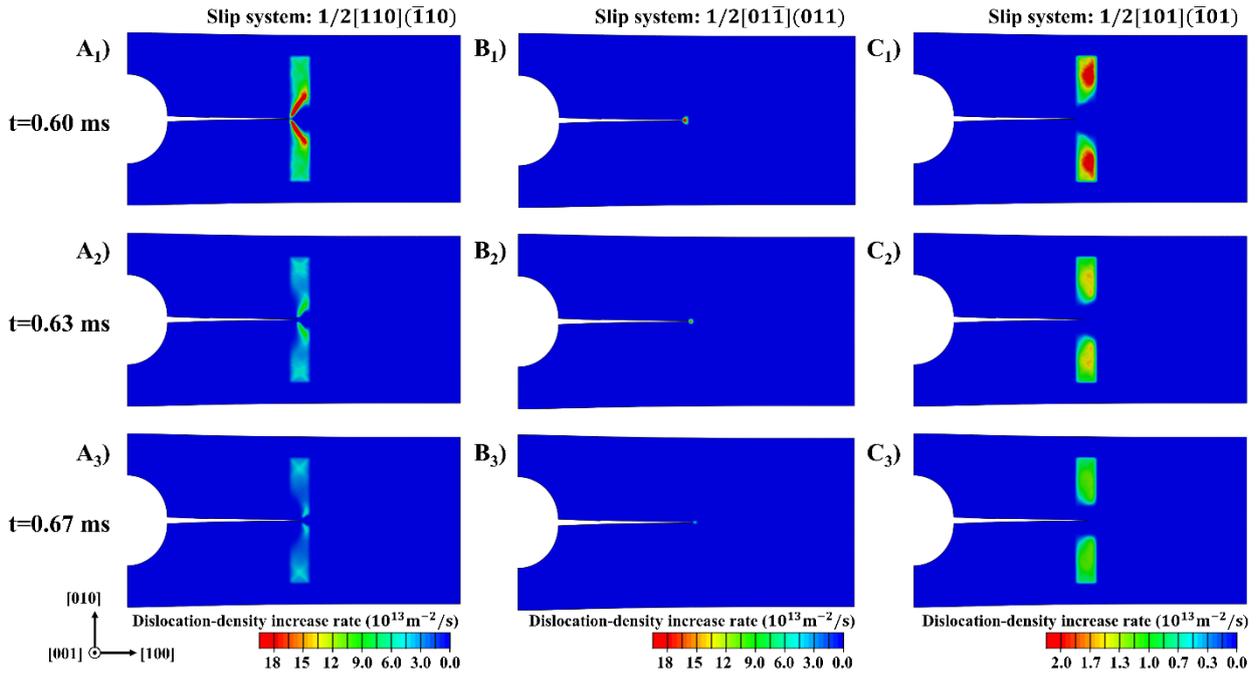

*Figure 5. Time evolution of dislocation density changing rate during the process of the crack-tip traversing the dislocation-rich region. $A_1$)-$A_3$) Results of $1/2[110](\bar{1}10)$ slip system. $B_1$)-$B_3$) results of $1/2[01\bar{1}](011)$ slip system. $C_1$)-$C_3$) Results of $1/2[101](\bar{1}01)$ slip system. Note that different color scales are used in $C_1$)-$C_3$).*

### 4.2.2. Crack-front shape during propagation

To capture the crack-front shape for the actual situation, a 3D model is established (**Fig. 6**). The dimensions of the 3D sample are width $W$ = 4 mm, height $H$ = 4 mm, thickness $L_t$ =1 mm, respectively, with a hole radius of $R$ = 0.6 mm. The initial crack length $a$ = 1 mm. The dislocation-rich region is located near the top surface, extending to a depth of 150 μm. These dimensions are set in the simulation to closely resemble the experimental design.

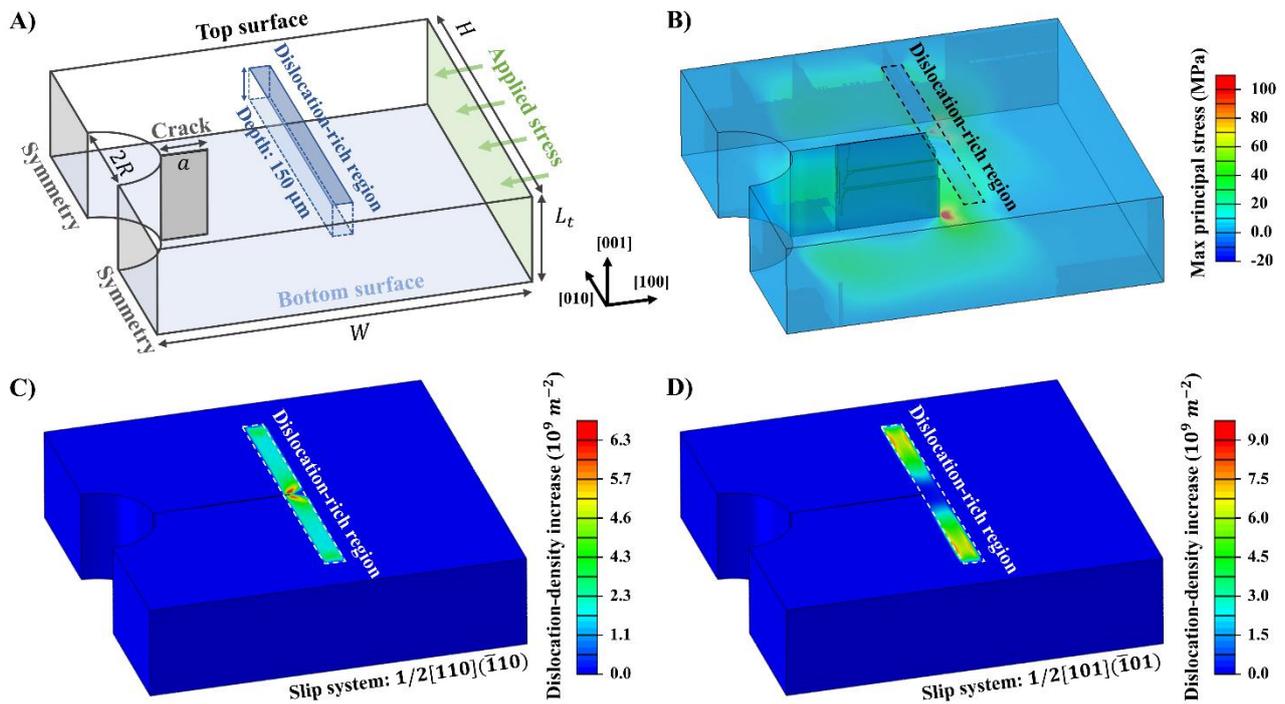

*Figure 6. A) Geometric dimensions corresponding to the 3D configuration: the width W = 4 mm, height H = 4 mm, and thickness $L_t$ = 1 mm, with a hole radius of R = 0.6 mm. The initial crack length is a = 1 mm. B) The distribution of maximum principal stress at the moment of crack interaction with the dislocation-rich region. C) and D) The distribution of dislocation-density increase in $1/2[110](\bar{1}10)$ and $1/2[101](\bar{1}01)$ slip system, respectively, corresponding to the same time as B.*

For the 3D case, the mechanism by which dislocations influence crack propagation remains largely consistent with the 2D scenario. Near the top surface, where the dislocation barrier is present, the crack propagation slows down while the crack in the bulk maintains its velocity (**Supplementary Video 2**). This results in a distinct 3D crack-front shape as outlined in **Fig. 7**. However, once the crack fully traverses the dislocation zone, this 3D characteristic diminishes as the crack accelerates and returns to a straighter crack front. These findings are consistent with the experimental observations in **Fig. 3**.

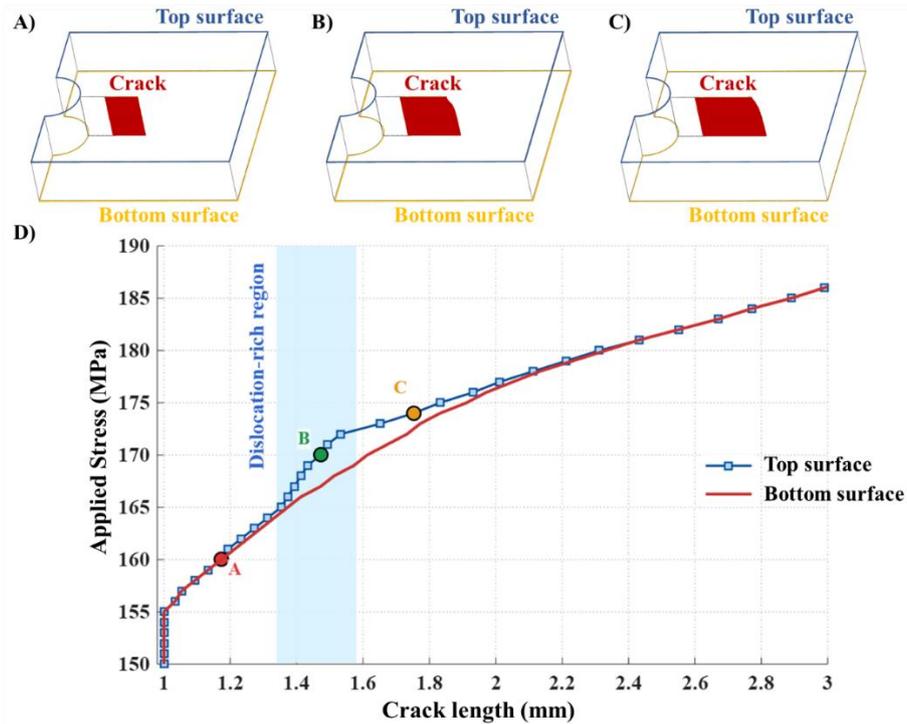

*Figure 7. A)-C) Sequential snapshots depicting the evolution of crack front morphology. D) Correlation between applied stress and resultant crack length, comparing the top surface (exhibiting dislocation shielding effect) with the bottom surface (lacking dislocation shielding).*

## 5. Discussion

### 5.1. Stable crack growth for *in situ* observation

The DCDC geometry allows for controlled crack propagation and arrest, as demonstrated above (**Figs. 1 & 3**) to capture the "snapshots" of crack tip – dislocation (plastic zone) interaction in consecutive steps. Note that in the experimental setup (see **Sec. 2.2**), having a copper interlayer combined with the finite machine stiffness resulted in a slow load decay (after the target load is reached, it drops with approximately 1 N/s), further aiding the already stable behaviour of the crack propagation. The mechanical drift from observation under load, however, does not allow at this stage to use magnifications above 1000x, meaning that a high-resolution tracking of individual dislocations in ECCI has not been reached yet. This remains a challenge for future studies.

While the sample has a thickness of ~2 mm, the dislocations introduced by Brinell indenter scratching only reach down to ~150 - 200 µm. The sample can be seen as a composite, with the dislocation barriers in the top 10% layer, and an unobstructed crack path below. Therefore, the observed surface portion of the crack is expected to be arrested in the barrier, while the sub-surface crack propagates further, as revealed by the 3D simulation in **Fig. 7**. In some cases where the top-surface crack was deflected and stopped in a dislocation-rich zone, the main crack resurfaced behind the dislocation barrier after a few minutes of constant load hold, which can be distinguished due to its perfectly straight shape (see **Fig. S1**). Decreasing the sample thickness to <500 µm, namely having a higher volume fraction of dislocation-rich zones, appears logical but would complicate the compression testing because of the lower stiffness and tendency for bending of the whole sample. As of now, there is no dislocation-engineering method for ceramics to reach significant homogeneous densities over ~$10^{14}$ m$^{-2}$ in a volumetric manner (bulk, uniform distribution). Nevertheless, our first attempt with this composite character is useful in providing new insight based on the observation in the near-surface region for crack propagation.

**5.2. Crack tip – Dislocation interaction**

Both experiment and CP-PFM modelling present a lower crack extension rate ($\frac{\partial a}{\partial \sigma_A}$) within the dislocation-rich barrier as compared to the reference, dislocation-sparse region in between. This means a deceleration of the crack propagation when entering the barrier, and an acceleration after breaking through the barrier. The main crack without barriers extending further than the top crack under the same load (see previous section and **Fig. 7**) also demonstrates this behaviour.

In **Fig. 3A**, the stress field around the crack tip is much smaller while passing the barrier, suggesting a lower local stress intensity factor $k_{I,local}$ in presence of the high-density dislocations. The influence of the stress field of the dislocations on the local stress intensity factor can be modelled using a shielding term $k_D$ [10]:

$$k_{I,local} = K_I + k_D \tag{11}$$

Each dislocation causes either a positive (anti-shielding) or negative (shielding, meaning toughening) contribution, which sums up to the total $k_D$. Hence, $k_D$ is a function of the Burgers vector and relative positions of all dislocations in the vicinity of the crack tip [10, 57]. According to Ref. [58], the applied load is a linear function of crack length. Therefore, the increment of applied load necessary for a unit crack extension can be expressed as:

$$\frac{d\sigma}{da} \propto F(H,R) K_{IC} \tag{12}$$

where $\sigma$ is the applied stress, $F(H,R)$ is a shape factor dependent on the specimen geometry. $K_{IC}$ is the critical stress intensity factor.

Based on **Eq. 12**, the slope of the curve in **Fig. 4B** can reflect the change of $K_{IC}$, which is indicative of the shielding effect by dislocations. This value is approximately increased by 4.7 times when dislocations are present, compared to the dislocation-free case. This indicates that the dislocation-rich region significantly enhances the apparent critical stress intensity factor, suggesting that plastic flow near the crack tip, rather than crack surface energy, becomes the dominant factor impeding crack advance. For the 3D case, the slope increases by roughly 2 times (**Fig. 7D**). The observed dislocation toughening effect is less pronounced than in the 2D scenario. This difference is attributed to the more advanced crack segment beneath the dislocation-rich region (**Fig. 7B-D**), which provides additional driving energy for the propagation of the top surface crack, thereby compromising the overall toughening.

As the DCDC method described here does not allow for high enough magnification for dislocation resolution, the active mechanism remains to be explored in the future. We discuss the following possible scenarios based on the dislocation structure of the scratch tracks:

1) The pre-engineered dislocations are of random Burgers vector within the activated slip systems at room temperature. Hence, except for very minor random local disturbances, half of the dislocations have shielding character $(k_{D,\ shielding} < 0)$, while the other half have anti-shielding character $(k_{D,\ anti-shielding} > 0)$. Therefore, $k_D$ would be 0 and there is no direct shielding from these dislocation structures.

2) A highly localized region with high-density dislocations serves as a center for dislocation multiplication. Dislocation glide caused by the crack-tip stress field consumes mechanical work. As every dislocation can become a dislocation source (especially when pinned), the pre-engineered dislocations could serve as multiplication centres, effectively lowering the emission stress. Emitted dislocations are always of shielding character. This "aided emission" is the most probable toughening mechanism (see **Fig. 5 and 7C-D**). Therefore, the engineered dislocations provide more potential dislocation sources that can contribute to plastic work at the crack tip.

The major slip system influenced by the crack-tip stress field is ½$[110](\bar{1}10)$, as well as the equivalent ½$[1\bar{1}0](110)$ slip system (mirrored along the crack plane). As the crack tip approaches the dislocation-rich region, significant dislocation glide is observed on the 45-degree plane intersecting the crack tip. The motion of these dislocations shields the crack-tip stress field, thereby requiring higher energy for the formation of new crack surfaces. This shielding effect also reduces the crack-propagation velocity.

For the coordinate system in **Fig. 4**, both the stress components $\sigma_{xx}$ and $\sigma_{yy}$ contribute to the resolved shear stress on slip system ½$[110](\bar{1}10)$. As the crack tip approaches the dislocation-rich region, significant dislocation glide is observed on the 45-degree plane intersecting the crack tip (see **Fig. 5A1-A3**). For the slip system ½$[011](01\bar{1})$, despite its Schmid factor induced by $\sigma_{xx}$ being zero under the considered loading conditions, the stress component $\sigma_{yy}$ provided by the crack-tip singular stress field can induce dislocation motion. Consequently, the dislocation-density increase rate is localized to a small region adjacent to the crack tip (see **Fig. 5B1-B3**). Additionally, dislocations activities of slip system ½$[101](10\bar{1})$ are mainly associated with deformation incompatibility at the edge of the dislocation-rich region. Their contributed plastic strain is one order of magnitude lower than those of the other two slip systems ½$[110](\bar{1}10)$ and ½$[101](10\bar{1})$. From an energy perspective, all plastic activities can dissipate energy, thereby reducing the energy available to drive crack propagation at the tip. As a result, the crack-propagation velocity decreases, and the required applied stress intensity factor increases.

On another note, MgO exhibits dislocation forest-hardening [53,59], meaning a high dislocation density (e.g. >$10^{14}$ m$^{-2}$) may also suppress the mobility of some dislocation. This can make it more difficult for the crack tip to plough through a dislocation cloud and move the dislocations, rendering crack-tip plasticity ineffective. Due to the back stress of the imprinted dislocations, the emission of dislocations could also be more difficult. This suggests that determining the optimal dislocation density to achieve the best toughening effect is a topic worthy of further in-depth investigation.

## 6. Conclusions

The double cleavage drilled compression (DCDC) geometry was used to achieve stable crack growth in single-crystal MgO. As stable crack growth allows to arrest the crack by simply holding (or slightly reducing) the load, electron channelling contrast imaging (ECCI) in a scanning electron microscope can be used for characterization at every increment of crack extension. By placing dislocation-rich barriers in the form of scratch tracks perpendicular to the crack's path, the ingress, passing and escape of the crack tip in such a zone can be observed quasi *in situ*. A coupled crystal plasticity and phase field model was employed to elucidate the experimental observations, where tuning the dislocation density allows for direct control of crack propagation in MgO. Under the considered conditions, a localized dislocation density on the order of ~1×$10^{15}$ m$^{-2}$ can reduce the crack growth rate by approximately 85% (modelling) or cracks are halted completely (experiment). This study demonstrates the feasibility of toughening ceramic materials even at room temperature through dislocation engineering. Future research should explore optimized dislocation configurations by tuning dislocation density and dislocation volume to achieve maximal toughening performance.

**Acknowledgement**

O.P. and J. R. thank the DFG for financial support (Grant No. 414179371). X. F. acknowledges the support by the European Union (ERC Starting Grant, Project MECERDIS, Grant No. 101076167). Views and opinions expressed are, however, those of the authors only and do not necessarily reflect those of the European Union or the European Research Council. Neither the European Union nor the granting authority can be held responsible for them. Y. C. acknowledges the support by the National Natural Science Foundation of China (No. 12222205). We thank Prof. K. Higashida from Kyushu University for helpful discussions on crack tip-dislocation interaction in oxides. We also thank D. Isaia and P. Breckner for setting up the testing stages and writing the software for the scratching method.

**Conflict of Interests:** The authors declare no conflict of interests.

**Supplementary Materials:**

1. Video1: stress evolution on surface (corresponds to **Figure 3**)

2. Video2: 3D crack front shape (corresponds to **Figure 4**)

3. Figure S1: image of resurfacing bottom crack, crack bridging and deflection (**Figure S1**)

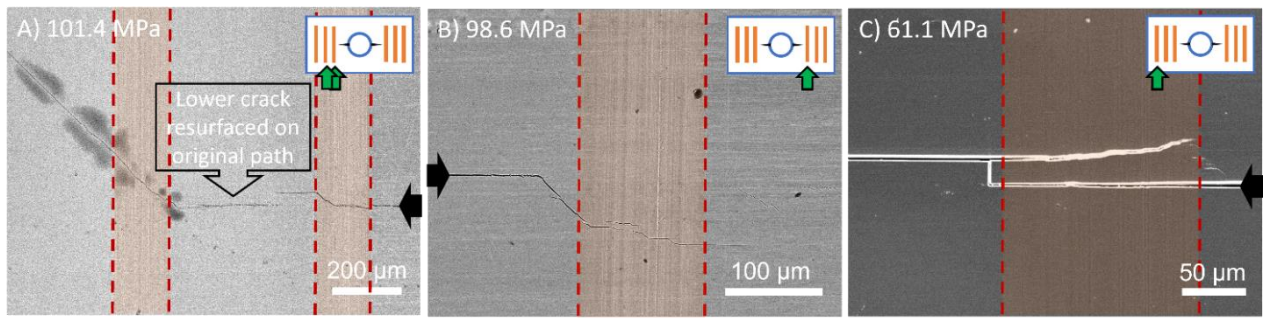

*Figure S1. Special Events captured during quasi in situ DCDC testing (all different samples). A) re-surfacing of bottom crack and crack deflection at the ingress of dislocation-rich zones, B) Crack deflection and bridging upon entering a dislocation-rich zone, C) crack deflection upon exiting a dislocation-rich zone.*